\documentclass[bm,cite,figures]{epl}

\usepackage{dcolumn}
\usepackage{amssymb}

\title{Direct transition to high-dimensional chaos through a global bifurcation}

\author{Diego Paz\'o\inst{1}\thanks{Present address: Max-Planck-Institut
f\"ur Physik komplexer Systeme - Dresden, Germany.\\ E-mail: \email{pazo@pks.mpg.de}}
 \and Manuel A. Mat\'{\i}as\inst{2}\thanks{E-mail: \email{manuel@imedea.uib.es}}}
\institute{
  \inst{1} Grupo de F\'{\i}sica Non Lineal, Facultade de
F\'{\i}sica \\ Universidade de Santiago de Compostela, E-15782
Santiago de Compostela, Spain\\
  \inst{2} Instituto Mediterr\'aneo de Estudios Avanzados,
IMEDEA (CSIC-UIB) \\ E-07122 Palma de Mallorca, Spain
}

\begin{document}

\maketitle

\begin{abstract}
In the present work we report on a genuine route by which 
a high-dimensional (with
$d\gtrsim 4$) chaotic attractor is created directly, i.e., without a
low-dimensional chaotic attractor as an intermediate step. 
The high-dimensional chaotic set
is created in a heteroclinic global bifurcation that yields 
an infinite number of unstable tori.
The mechanism is illustrated
using a system constructed by coupling three Lorenz oscillators. So,
the route presented here can be considered a prototype for high-dimensional
chaotic behavior just as the Lorenz model is for low-dimensional chaos.

\end{abstract}

\section{Introduction}
The field of Nonlinear Dynamics has achieved a state of maturity 
in the study and characterization of the transitions exhibited by
low-dimensional dissipative dynamical systems~\cite{Ott}.
In particular, transitions to chaotic behavior in these systems appear to
take place through a few well-known routes (or scenarios):
the period doubling cascade, the intermittency route, the routes involving the
destruction of quasiperiodic tori or the crisis route (see, e.g., \cite{Ott,Eckmann81,berge}
for a survey). Another possibility is that chaos appears through a global connection
to a fixed point \cite{wiggins}, e.g. Shil'nikov or Lorenz chaos~\cite{Sparrow}.
Characterizing these scenarios is 
important because they present universal features,
independently of the physical system involved. 

Contrastingly, not many studies have been published about the
transitions to chaotic attractors with dimension $d>3$, i.e., high-dimensional chaos.
Putting aside routes starting from a low-dimensional chaotic attractor (for instance
in the transition to hyperchaos where a second Lyapunov exponent becomes 
positive~\cite{Harrison99,Kapitaniak00}),
only a few cases have been reported in the literature. These 
involve generalizations of
low-dimensional routes to chaos through an extra oscillation: 
period doubling cascade of a torus \cite{Yang00} (instead of a limit cycle) or 
a Lorenz-type homoclinic connection to a limit cycle~\cite{moon97} (instead of a fixed point).
In these two cases the dimension of the high-dimensional chaotic attractor is (roughly)
{\it one\/} unit above the chaotic attractor arising from the corresponding 
low-dimensional route. 
In this framework, the following relevant question arises:
does it exist a {\em bona fide} direct transition to H-D chaos?
i.e. a mechanism that cannot be reduced to a low-dimensional route
with an extra oscillation?
(unlike in \cite{moon97,Yang00}).

High-dimensional ($d>3$) chaotic attractors {\it live\/} in spaces of 
large dimension $d\ge 4$, and it is, therefore, 
plausible the existence of routes to (H-D) chaos that cannot be envisaged
from low-dimensionality. 
Strictly one should be speaking here of {\it intermediate-dimensional\/} 
rather than high-dimensional attractors, as often in the literature
the dimension is by far larger than three. 
For instance, (very) high-dimensional attractors are usually related
to the regime of spatio-temporal chaos in representative nonlinear
partial differential equations (like the Kuramoto-Sivashinsky equation
\cite{bohr}).
The aspect that marks an important difference here is that these
regimes have been characterized mostly using statistical techniques~\cite{politi},
while here
we are proposing, instead, a geometric analysis, for which the
detailed study of phase spaces of dimension, say, $4$ and $5$ is already quite a 
challenge.
So, in this sense we present the first example of a genuine route
to H-D chaos, allowed by the
high enough dimensionality of the embedding space.

\section{System and overall picture}
The model studied here consists of three 
Lorenz oscillators~\cite{Sparrow,lorenz63} coupled unidirectionally. 
The evolution equations (an autonomous $9$-dimensional dynamical system) 
read:
\begin{eqnarray}
\left.
\begin{array}{rcl}
\dot{x_j}&=&\sigma(y_j-x_j)\\
\dot{y_j}&=&R\,\underline{x_j}-y_j-x_j\,z_j\\
\dot{z_j}&=&x_j\,y_j-b\,\,z_j
\end{array}
\right\} \quad j=1,\ldots,N=3 \label{eqlor}\ ,
\end{eqnarray}
where $\underline{x_j}=x_{j-1}$ for $j\neq 1$, introduces the
coupling and periodic boundary conditions 
are used: $\underline{x_1}=x_3$.

The study of~(\ref{eqlor})
has been suggested by the results of
the  experiment with three coupled Lorenz oscillators modeled by 
electronic circuits~\cite{sanchezpre98,sanchezijbc99}.
By increasing $R$, synchronized chaos among the three oscillators gives rise to
high-dimensional chaos (a Chaotic Rotating Wave, CRW~\cite{matiasepl97,sanchezijbc99}) and, 
finally, quasiperiodic and periodic behaviors. Here we focus on
the transition between `order' and H-D chaos, obtained when
$R$ is decreased.

\begin{figure}
\onefigure[width=4.in]{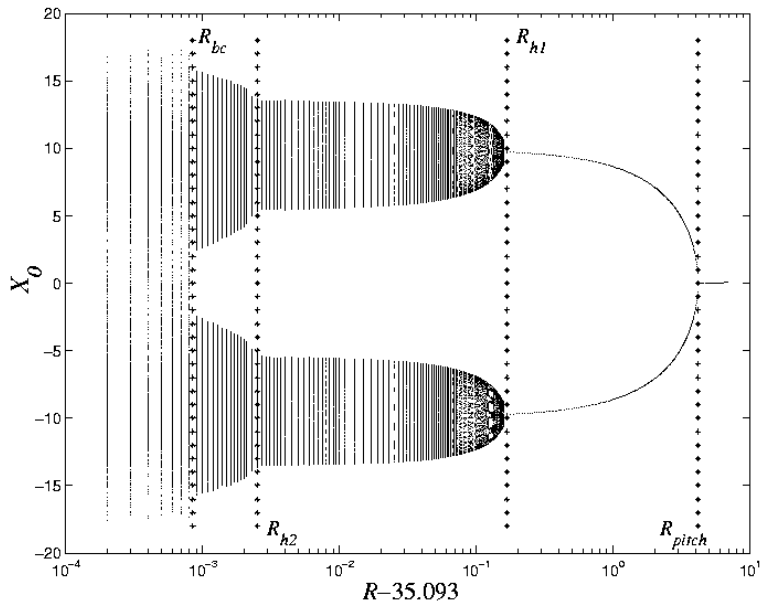} 
\caption[]{Bifurcation diagram representing the $X_0$ coordinates of the 
intersection with the Poincar{\'e} section (see text) as a function of the parameter $R$.  
The logarithmic scale has been adopted in the $x$-axis to better
resolve attractors existing in quite different ranges. Parameters in 
Eq.~(\ref{eqlor}): $\sigma=20$, $b=3$.}\label{diagram2}
\end{figure}

A high-frequency periodic
component (a rotating wave into the ring) is present along the route. Hence, 
to reduce the dimensionality of the problem, most 
of our presentation is based on 
an appropriate Poincar\'e section: $Im(X_1)=0$, $Im(\dot{X}_1)>0$ 
(we use the discrete Fourier modes, suggested by the circulant symmetry of the system, defined as 
$X_k = {1/N} \sum x_j \exp[{{2\pi i (j-1) k}/N}]$; analogously for $y$ and $z$ coordinates).
Figure~\ref{diagram2} depicts the coordinates $X_0$ of the intersections of 
the trajectories into the different attractors.
For large $R$, the system exhibits a 
Periodic Rotating Wave (PRW) \cite{matiasprl97,sanchezpre98,sanchezijbc99}, which
is a periodic motion with $2 \pi / 3$ phase difference between adjacent oscillators.
This state exhibits a pitchfork bifurcation at $R_{pitch}\approx 39.25$, in
which the single PRW attractor leads to two twin asymmetric PRW attractors.
By further decreasing $R$, the system exhibits two consecutive Hopf 
bifurcations at $R_{h1}\approx 35.26$ and 
$R_{h2}\approx 35.0955$ giving rise to two stable 
symmetry related 3D-tori~\cite{pazoijbc01}.
The robust existence of $\mathbb{T}^3$ attractors 
as well as the absence of appreciable
frequency lockings in the $\mathbb{T}^2$ attractors stems
from: 1) The disparity of the frequencies, such that resonances 
correspond to rationals with large denominators (and therefore narrow
Arnol'd tongues \cite{Kuznetsov});
2) The cyclic symmetry of the system:
the absence of lockings is a
general feature of modulated rotating waves (like our $\mathbb{T}^2$) in
systems with rotational symmetry \cite{rand}, so although this result
only holds exactly in the continuum limit, one expects some inhibition
of the lockings.

The H-D chaotic attractor appears 
at $R_{bc} \approx 35.09384$, and as may be seen in Fig.~\ref{diagram2}
the dynamics recovers the reflection symmetry lost in the pitchfork bifurcation.
The dimension of the chaotic attractor can be estimated
by means of a direct calculation of the correlation 
dimension~\cite{grassberger} that yields $D_2=3.96 \pm 0.05$, 
for $R=35.05$. Therefore, we must envisage a route that is able to create
a chaotic attractor with dimension around four. Taking this 
into account we list some evidences that allowed us
to understand the mechanisms involved in the creation of the H-D
chaotic attractor: 

{\em 1.} We observe that the average chaotic transient diverges at $R_{bc}$. 
The divergence follows a power law,
typical for boundary crises~\cite{grebogi87}, that convert 
chaotic attractors into chaotic transients:
$\langle\tau\rangle \sim (R-R_{bc})^\gamma$, $\gamma=-1.53\pm 0.06$.  

{\em 2.} The $\mathbb{T}^3$ attractors disappear 
at twin saddle-node bifurcations~\cite{footnote1}
at $R_{sn} \approx 35.09367$.
Therefore, as $R_{sn}<R_{bc}$, 
there exists a small range of {\em coexistence} between the $3D$-tori 
and the H-D chaotic attractor. Then, we conclude that the $\mathbb{T}^3$ attractors 
are not involved in the birth of the chaotic attractor.

{\em 3.} A value $R_{expl}\approx 35.11$ is found to define a transition below which
there are chaotic transients (in which, for some initial conditions, the 
trajectory 
approaches both $\mathbb{T}^2$ attractors, in a non-periodic manner, before being 
eventually attracted by one of them).
We observed that this point coincides (at least approximately) with a 
reordering of the unstable manifolds of the PRWs. 

\begin{figure}
\onefigure[width=1.0\textwidth]{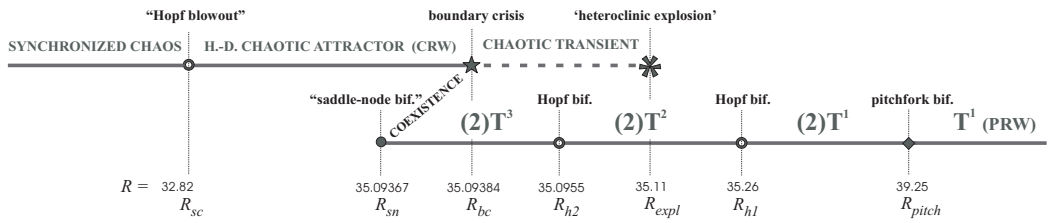}
\caption[]{Diagram representing schematically the transitions from
synchronized chaos (left) to a PRW (right).} 
\label{diagram1}
\end{figure}

\section{Theoretical analysis}

These evidences lead us to conclude that the transition
from periodic to chaotic behavior occurs following
the set of bifurcations shown in Fig.~\ref{diagram1}. 
It includes the creation
of the H-D chaotic set in a `heteroclinic explosion'
followed by its conversion
into an attractor through a boundary crisis.

\begin{figure}
\onefigure[width=5.2in]{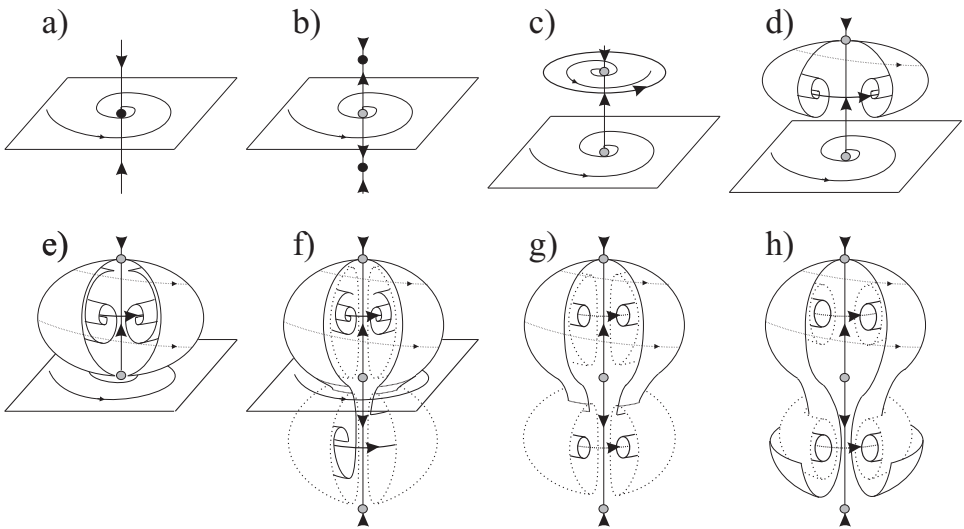}
\caption[]{Representation (a Poincar\'e section projected on $\mathbb{R}^3$) 
of the proposed heteroclinic route to create the high-dimensional chaotic
attractor.
Black and gray points correspond to stable and unstable
fixed points (cycles in the global phase space), respectively.} 
\label{3dscheme}
\end{figure}

Figure~\ref{3dscheme} illustrates geometrically the route to H-D chaos 
(note that we consider a Poincar\'e section that reduces the dimensions of
all the attractors by one):
the centered PRW (a)
becomes unstable through a pitchfork bifurcation (a$\rightarrow$b)
and two symmetry related PRWs appear (b). At a supercritical Hopf
bifurcation (b$\rightarrow$c) the 2D-tori appear. When $R$
is further decreased the 2D-tori become focus-type, and
as a result, the unstable manifold of the
asymmetric PRW forms a ``whirlpool'' \cite{shilnikov95} when approaching the
$\mathbb{T}^2$ (d). At $R_{expl}$ a double heteroclinic connection
between the asymmetric PRWs and the symmetric one occurs (e). At
this point the chaotic set, with a dense set
of unstable 3D-tori, and thus a dimension above 4, is created. In (f) the two simplest unstable 3D-tori are
represented with dotted lines; because of the heteroclinic birth
one of the frequencies of these tori is very small (formally zero at $R_{expl}$). 
Notice that the plot shows 
several forbidden intersections of manifolds
that are
are unavoidable because of the projection onto
$\mathbb{R}^3$ (analogous `visual' effect occurs when the Lorenz system
is projected onto $\mathbb{R}^2$, $x-z$ plane). 
Twin secondary Hopf bifurcations (f$\rightarrow$g) render unstable the
2D-tori and give rise to two stable 3D-tori (g). For smaller $R$,
the unstable manifolds of the asymmetric PRWs
do not connect to the stable 3D-tori (h), and
the chaotic set becomes attracting. This last step (g$\rightarrow$h) is analogous
to the boundary crisis occurring in the 
Lorenz system~\cite{grebogi87}.  
Finally, it is to be stressed that the  
reflection symmetry plays a fundamental role; whereas
the existence of stable $\mathbb{T}^3$ is not needed 
[focus type $\mathbb{T}^2$ suffice to induce whirlpools that
finally become 2D heteroclinic connections, 
Fig.~\ref{3dscheme}(d,e)].

\section{Lyapunov spectrum}

\begin{figure}
\onefigure[width=4in]{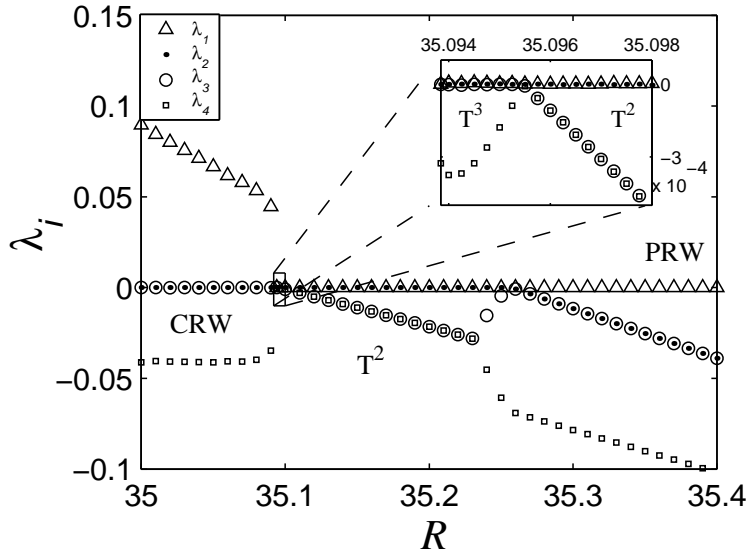}
\caption[]{The four largest Lyapunov exponents
($\lambda_{i=1-4}$) as a function of $R$. Four
regions can be distinguished according to their spectra. 
Note that in the chaotic region we obtain
$\lambda_2 \simeq \lambda_3 \simeq 0$ and $\lambda_1 \gtrsim |\lambda_4|$ which
implies, according to the Kaplan-Yorke conjecture, an information
dimension $D_1\gtrsim4$.} 
\label{lyap_tran}
\end{figure}

Let us see now how the route shown above relates to the Lyapunov 
spectrum. Figure~\ref{lyap_tran} shows the leading
Lyapunov exponents (LEs) in the transition from periodic 
behavior to chaos. 
Please, notice that similar Lyapunov spectrum 
has been obtained considering $N=4$ and a more general 
coupling (cf. Fig. 5(a) in \cite{Yang00}). 
In the chaotic region only one exponent
is positive whereas two vanish and, one of the  
negative LEs is close to zero. It is remarkable that $\lambda_1 \gtrsim |\lambda_4|$, 
which implies $D_1 \gtrsim 4$, according to the Kaplan-Yorke 
conjecture~\cite{footnote0}


\begin{table}
\caption{The five smallest Lyapunov exponents for both, chaotic (CRW) and 
three-frequency quasiperiodic ($\mathbb{T}^3$), attractors 
coexisting at $R=35.0938$.}
\label{tabla}
\begin{center}
\begin{tabular}{c c c} \hline\hline
   &  $\lambda_i$(CRW) &  $\lambda_i$($\mathbb{T}^3$) \\ \hline
  $\quad\lambda_5=\lambda_6\quad$ & $\quad -5.255 \quad$& $\quad -5.203 \quad$ \\
  $\lambda_7=\lambda_8$ & $\quad-18.612 \quad$ & $\quad-18.652 \quad$\\
  $\lambda_9$ & $\quad-24.273 \quad$ & $\quad-24.290 \quad$\\ \hline \hline 
\end{tabular}
\end{center}
\end{table}

Also, the fifth to ninth (non-leading) LEs for the chaotic attractor and 
the twin 3D-tori are very similar (see table~\ref{tabla}) which suggests that the 3D-tori and
the H-D chaotic attractor ``live'' in the same four-dimensional subspace
(as we have implicitly assumed above). And thus,
thinking in terms of some kind of generalization of the 
Birman-Williams theorem~\cite{gilmore}, the
template of the chaotic attractor can be visualized as a 4D 
branched manifold. In the same way, the (butterfly) Lorenz attractor
may be understood as a two-dimensional branched manifold, 
with a `tear point' at the origin~\cite{footnote2}.

\section{Further remarks and conclusions}

Some further comments are in order with respect to the relation 
between the LEs and the geometric sketch depicted in Fig.~\ref{3dscheme}.
It shows a 
mechanism, through heteroclinic connections, 
to create a chaotic attractor 
containing an infinity of unstable tori (UT)  ---recall that only
the two simplest ones are drawn--- instead of unstable periodic orbits (UPOs). 
If, as it occurs in our system,
there is an additional frequency 3D-UT are created instead.   
In consequence, there are three neutral directions and 
the chaotic attractor should have three null LEs. 
Fig.~\ref{lyap_tran} shows instead two vanishing
and one slightly negative LEs. 
This shift is a consequence of the fact that 
a generic perturbation on the mechanism shown in
Fig.~\ref{3dscheme} will destroy its symmetry (and consequently its simplicity).
In analogy to previous works~\cite{gaspard,kirk91} dealing with the effect of
non-symmetric terms on the normal form of codimension-two points, 
we expect (generic) homoclinic connections to replace heteroclinic 
connections. 
A double (`figure-eight') homoclinic to the symmetric PRW as well
as homoclinic connections to the asymmetric 
PRWs will occur~\cite{footnote3}
Consequently, in the perturbed scenario an infinity of UT (instead
of 3D-UT) are created. This explains the absence of a third vanishing Lyapunov
exponent. 
But it is important to emphasize
that as long as the exact mechanism is closely related to the one shown
in Fig.~\ref{3dscheme}, the largest negative LE is
close enough to zero ($\left|\lambda_4\right|< \lambda_1$) 
to get an information dimension above {\em four}. 

In this Letter, we have reported 
the creation of a high-dimensional chaotic (but not hyperchaotic) 
attractor without intermediate low-dimensional chaos. 
The structure of the global bifurcations,
underlying this route, 
leads to the emergence of 
a chaotic attractor with dimension $D_1 \gtrsim 4$ 
(or $D_1 \gtrsim 3$ if the fast rotating wave present all along the
route is considered to increase trivially the dimension in one unit).
A characteristic of the emerging chaotic attractor is the presence of a very low
frequency component, reminiscent of the heteroclinic birth of the 
chaotic 
set (this may have been observed in fluid convection
experiments~\cite{dubois}).
We believe that the knowledge of the 
direct routes to high-dimensional chaos 
may lead to a re-thinking of previous experimental results,
as well to understand and design future experiments.

\acknowledgments
DP thanks Dr.\ M.\ Zaks for introducing him
to the computation of UPOs. This work was supported by MEC (Spain) and FEDER under
Grants  BFM2001-0341-C02-02, FIS2004-00953, and FIS2004-05073-C04-03. DP acknowledges the financial
support by  {\em Secretar\'{\i}a  Xeral 
de Investigaci\'on e Desenvolvemento} of the 
{\em Xunta de Galicia}.
We acknowledge support from the MPIPKS (Dresden).



\begin{thebibliography}{10}
\bibitem{footnote1}
{After the 3-D tori 
disappear no attractor remains at their place.
Further, we observe the fourth Lyapunov exponent 
$\lambda_4 \propto -(R-R_{sn})^{1/2}$,
whereas $\lambda_{1-3} \simeq 0$. 
Thus, it can be assumed that the
$\mathbb{T}^3$ attractors disappear through saddle-node bifurcations,
although a deviation of this mechanism cannot be excluded in a
very small interval of $R$.}
\bibitem{footnote0}
{The Kaplan-Yorke conjecture \cite{kapyor} 
states that 
$D_1 = K+\sum_{j=1}^K (\lambda_j/\vert
\lambda_{K+1}\vert)$, being $K$ the largest integer such that
$\sum_{j=1}^K \lambda_j\ge 0$, where the $\lambda_j$ are the
Lyapunov exponents ordered from larger to smaller.}
\bibitem{footnote2}
{A simplified vision of the dynamics 
inside an attractor is to consider that the trajectories inside
are contained in a branched manifold. For the Lorenz attractor, this one consists of 
two  planar sectors, 
the two lobes of the attractor, that close to the origin are `glued' in a single 
surface (and trajectories coming from both lobes are no longer distinguishable).
Thus, the dynamics is purely 2-dimensional, except at the jumps between
both lobes where an extra dimension is needed.}
\bibitem{footnote3}
{This kind of bifurcations produce quasi-attractors
that exhibit infinitely many bifurcations of various types and cannot be described completely 
~\cite{gonchenko}.}
\end{thebibliography}

\end{document}